%% file: hydrodynamics-Weyl-short-v12.tex
\documentclass[aps,prl,superscriptaddress,twocolumn,10pt]{revtex4-1}
\usepackage{epsfig}
\usepackage{graphicx}
\usepackage{amsmath}
\usepackage{bm}
\usepackage{amssymb}
\usepackage{color}

\definecolor{purple}{rgb}{0.8,0,0.6}
\definecolor{darkgreen}{rgb}{0.00,0.6,0.00}
\def\mytitle{Consistent hydrodynamic theory of chiral electrons in Weyl semimetals}

\begin{document}

\title{\mytitle}
\date{December 25, 2018}
%\date{\today}

\author{E.~V.~Gorbar}
%\email{gorbar@bitp.kiev.ua}
\affiliation{Department of Physics, Taras Shevchenko National Kiev University, Kiev, 03680, Ukraine}
\affiliation{Bogolyubov Institute for Theoretical Physics, Kiev, 03680, Ukraine}

\author{V.~A.~Miransky}
%\email{vmiransk@uwo.ca}
\affiliation{Department of Applied Mathematics, Western University, London, Ontario, Canada N6A 5B7}

\author{I.~A.~Shovkovy}
%\email{igor.shovkovy@asu.edu}
\affiliation{College of Integrative Sciences and Arts, Arizona State University, Mesa, Arizona 85212, USA}
\affiliation{Department of Physics, Arizona State University, Tempe, Arizona 85287, USA}

\author{P.~O.~Sukhachov}
%\email{psukhach@uwo.ca}
\affiliation{Department of Applied Mathematics, Western University, London, Ontario, Canada N6A 5B7}

\begin{abstract}
The complete set of Maxwell's and hydrodynamic equations for the chiral electrons
in Weyl semimetals is presented. The formulation of the Euler equation takes into account the
explicit breaking of the Galilean invariance by the ion lattice. It is shown that the Chern--Simons
(or Bardeen--Zumino) contributions should be added to the electric current and charge densities
in Maxwell's equations that provide the information on the separation of Weyl nodes in
energy and momentum. On the other hand, these topological contributions do not directly affect the Euler equation and the energy conservation
relation for the electron fluid. By making use of the proposed \emph{consistent hydrodynamic} framework, we
show that the Chern--Simons contributions strongly modify the dispersion relations of collective
modes in Weyl semimetals. This is reflected, in particular, in the existence of distinctive
{\em anomalous Hall waves}, which are sustained by the local anomalous
Hall currents.
\end{abstract}

\maketitle

{\em Introduction.---}
The formulation of relativistic hydrodynamics was proposed more than half a century ago (see,
for example, Ref.~\cite{Landau:t6}). It is widely used in various fields of physics, ranging
from nuclear physics to astrophysics and cosmology. In interacting systems close to equilibrium,
the hydrodynamic equations describe the space-time evolution of conserved quantities (e.g.,
energy, momentum, electric charge, etc.) in the limit of long wavelengths and
long time scales. Recently, relativistic hydrodynamics was also generalized to plasmas
made of chiral fermions \cite{Son:2009tf,Sadofyev:2010pr,Neiman:2010zi,Hidaka:2017auj} where the chiral
charge is included as an additional degree of freedom, whose conservation is violated only
by the chiral anomaly.

The potential relevance of hydrodynamics for the electron transport in solids is an old
idea too \cite{Gurzhi}. However, it can be realized only when the electron-electron scattering
rate is much larger than the rate of electron scattering on phonons and impurities. Experimentally,
the hydrodynamic transport of relativistic-like electrons was recently
observed in graphene \cite{Crossno,Ghahari}, which is a two-dimensional Dirac material.

It is reasonable to expect that the hydrodynamics is also relevant for the electron fluid in
Dirac and
Weyl semimetals. As in graphene, their low-energy quasiparticles are relativistic-like
fermions, although propagating in the three rather than two dimensions.
Moreover, the time-reversal (TR) symmetry and/or the parity inversion (PI) are broken in Weyl semimetals.
As a result, the Weyl nodes come in pairs of opposite chirality that are separated
in momentum and/or energy. The corresponding distances are quantified by the chiral shift $\mathbf{b}$ and parameter $b_0$, respectively.
(For recent reviews of Weyl semimetals, see Refs.~\cite{Yan-Felser:2017-Rev,Hasan-Huang:2017-Rev,Armitage-Vishwanath:2017-Rev}.)

The experimental confirmation of the hydrodynamic regime in the Weyl semimetal tungsten diphosphide ($\mbox{WP}_2$) was recently reported in Ref.~\cite{Gooth:2017}, where the dependence of the electrical
resistivity on the channel width provides compelling evidence for the hydrodynamic
transport. This interpretation is further supported by the observation of the Wiedemann--Franz law violation with the lowest value of the Lorenz number ever reported.

Previously, the equations of relativistic hydrodynamics were used to describe the negative magnetoresistance
\cite{Landsteiner:2014vua,Lucas:2016omy} and the thermoelectric transport \cite{Lucas:2016omy}
in Weyl semimetals. However, the corresponding approach lacks important information regarding
the separation of the Weyl nodes in energy and momentum. The situation is similar to the
conventional chiral kinetic theory \cite{Son,Stephanov:2012ki}
in which this information is also missing. In addition, such a kinetic theory suffers from an acute problem of the local
nonconservation of the electric charge when both electromagnetic and strain-induced pseudoelectromagnetic fields
are applied to the system \cite{Pikulin}. We resolved this problem
in Ref.~\cite{Gorbar:2016ygi} by amending the electric charge and current densities with the Chern--Simons contributions
\cite{Landsteiner:2013sja,Landsteiner:2016}. The latter are also known as the Bardeen--Zumino terms connected with the consistent anomaly in high
energy physics \cite{Bardeen}.
In this paper, by using the consistent chiral kinetic theory \cite{Gorbar:2016ygi} in the relaxation time approximation,
we derive the hydrodynamic equations and
study the implications of the Chern--Simons contributions on the properties of the collective excitations of the electron fluid in
Weyl semimetals.

{\em Origin of the Chern--Simons contributions.---}
In order to demonstrate the need for the Chern--Simons contributions in the hydrodynamic
theory of the electrons in Weyl materials, let us consider a two-band model of a Weyl semimetal
defined by the following Hamiltonian \cite{Ran,Franz:2013}:
\begin{equation}
\label{BZ-lattice-H-def}
\mathcal{H}_{\rm latt} =t_0\sin{(ak_z)} +\sum_{i=x,y,z}\sigma_id_i(\mathbf{k}),
\end{equation}
where the first term $t_0\sin{(ak_z)}$ is responsible for breaking the PI symmetry,
$\bm{\sigma}=(\sigma_x,\sigma_y,\sigma_z)$ are the Pauli matrices, and $\mathbf{d}(\mathbf{k})$
is a periodic function of the quasi-momentum $\mathbf{k}$. A simple Weyl
semimetal with a broken TR symmetry is defined by $d_{x} = \Lambda \sin{(ak_x)}$,
$d_{y} = \Lambda \sin{(ak_y)}$, and $d_{z} =t_0 +t_1\sum_{i=x,y,z}\cos{(ak_i)}$, where $a$ denotes the
lattice spacing, whereas $\Lambda$, $t_0$, and $t_1$ are material-dependent parameters.
In this model, the opposite-chirality Weyl nodes are separated by $2eb_z/(c\hbar)$ in momentum
and by $2eb_0$ in energy, where $b_0 = t_0\sin{(eab_z/c\hbar)}/e$, $b_z=c\hbar \arccos{\left[-\left(t_0+2t_1\right)/t_1\right]}/(ea)$,
and $e$ is the absolute value of the electron charge.

The topological features of Weyl semimetals are captured by a nontrivial Berry
curvature \cite{Berry:1984}, i.e.,
\begin{equation}
\Omega_{i}=\sum_{l,m=x,y,z}\frac{\epsilon_{i l m}}{4\hbar|\mathbf{d}|^3}
\left(\mathbf{d}\cdot\Big[\partial_{k_{l}}\mathbf{d}
\times\partial_{k_{m}}\mathbf{d}\Big]\right).
\label{BZ-lattice-topology-inv-Berry}
\end{equation}
This result is valid in the whole Brillouin zone and, unlike its simplified monopole analogs in
linearized low-energy models, it captures the nontrivial dependence
on $b_0$ and $\mathbf{b}$.
By making use of the chiral kinetic theory approach \cite{Son:2012wh}, in the limit of zero
temperature ($T=0$) and at the neutrality point
(with the vanishing electric and chiral chemical potentials $\mu=\mu_5=0$), we find the following topological charge and current densities
determined by the Berry curvature:
\begin{eqnarray}
\label{BZ-charge}
\rho_{\text{{\tiny CS}}} &=& -\frac{e^2}{c}\int\frac{d^3k}{(2\pi)^3} (\mathbf{B}\cdot\mathbf{\Omega})
= -\frac{e^3}{2\pi^2\hbar^2c^2}\,(\mathbf{b}\cdot\mathbf{B})\,,\\
\label{BZ-current}
\mathbf{j}_{\text{{\tiny CS}}} &=&-e^2\int\frac{d^3k}{(2\pi)^3}\,
\left\{\frac{(\mathbf{v}\cdot\mathbf{\Omega}) \mathbf{B}}{c} + \left[\mathbf{E}\times
\mathbf{\Omega}\right]\right\} \nonumber\\
&=& -\frac{e^3}{2\pi^2\hbar^2c}\,b_0 \mathbf{B}
+ \frac{e^3}{2\pi^2\hbar^2c}\,[\mathbf{b}\times\mathbf{E}],
\end{eqnarray}
where $\mathbf{E}$ and $\mathbf{B}$ are the electric and magnetic fields, respectively, and $\mathbf{v}=\hbar^{-1}\left[at_0\cos{(ak_z)}\hat{\mathbf{z}}+\partial_{\mathbf{k}}|\mathbf{d}|\right]$. (Here, $\hat{\mathbf{z}}$ denotes the unit vector in the $+z$ direction.)
The topological nature of these terms as well as the absence of the ``matter" contributions is evident from the lack of the distribution function in Eqs.~(\ref{BZ-charge}) and (\ref{BZ-current}).
The Chern--Simons contributions $\rho_{\text{{\tiny CS}}}$ and $\mathbf{j}_{\text{{\tiny CS}}}$ play an essential role in the consistent chiral kinetic theory
\cite{Gorbar:2016ygi}. As is easy to check, they originate primarily from the filled electron
states deep below the Fermi surface. This explains why such contributions to the
charge and current densities are usually missing in linearized semiclassical low-energy models, such as the chiral kinetic theory.

In this connection, we should also emphasize that the expressions
for $\rho_{\text{{\tiny CS}}}$ and $\mathbf{j}_{\text{{\tiny CS}}}$ are proportional to a winding number
of the mapping of a two-dimensional section of the Brillouin zone onto the unit sphere \cite{BZ-lattice}.
This fact accounts for the topological nature of $\rho_{\text{{\tiny CS}}}$ and $\mathbf{j}_{\text{{\tiny CS}}}$
and explains their robustness.

{\em Hydrodynamic equations.---}
In order to derive the hydrodynamic equations for the chiral electron fluid in Weyl semimetals,
we utilize the consistent chiral kinetic theory \cite{Gorbar:2016ygi} in the relaxation-time approximation.
By following the standard approach \cite{Landau:t10,Huang-book}, the Euler
equation and the energy conservation relation are obtained by multiplying
the corresponding kinetic equation with the quasiparticle momentum and energy, respectively, and integrating over the momentum
(for details, see the Supplemental Material).
As expected, many terms of the resulting hydrodynamic equations agree with those in
Refs.~\cite{Son:2009tf,Sadofyev:2010pr,Neiman:2010zi,Landsteiner:2014vua,Lucas:2016omy}.
In order to highlight the distinctive features
of our framework, here we present the abbreviated version of the equations in which the terms with
spatial derivatives are omitted, i.e.,
\begin{eqnarray}
\label{MHD-Euler}
&&\frac{1}{v_F}\partial_t \left(\frac{\epsilon+P}{v_F} \mathbf{u} + \sigma^{(\epsilon,B)}\mathbf{B}\right) = -en\left(
\mathbf{E} +\frac{1}{c}\left[\mathbf{u} \times \mathbf{B} \right]\right)\nonumber\\
&&+\frac{\sigma^{(B)} (\mathbf{E}\cdot\mathbf{B})}{3v_F^2} \mathbf{u}  -\frac{\epsilon+P}{\tau v_F^2} \mathbf{u}
+ O(\bm{\nabla}_{\mathbf{r}})
\end{eqnarray}
and
\begin{equation}
\label{MHD-energy}
\partial_t \epsilon = -\mathbf{E}\cdot \left(en\mathbf{u}-\sigma^{(B)}\mathbf{B}\right)
+ O(\nabla_{\mathbf{r}}).
\end{equation}
(The full expressions including also the vorticity effects, such as the chiral vortical effect \cite{Chen:2014cla} and the terms with magnetovorticity coupling \cite{Hattori:2016njk}, are given in the Supplemental Material.)
In these equations, $\epsilon$ and $P$ are the electron energy density and pressure, respectively,
$n$ is the electron number density, $\mathbf{u}$ is the electron fluid velocity, $\tau$ is the relaxation
time, and $v_F$ is the Fermi velocity.
The effects of the chiral anomaly including the chiral magnetic effect current
\cite{Kharzeev:2007tn,Kharzeev:2007jp,Fukushima:2008xe}, as well as the energy-momentum flow are captured in the hydrodynamic equations by the terms $\propto \sigma^{(B)}\mathbf{B}$ and
$\propto \sigma^{(\epsilon, B)}\mathbf{B}$, respectively, with $\sigma^{(B)} = e^2\mu_5/(2\pi^2\hbar^2c)$ and $\sigma^{(\epsilon, B)} = -e\mu\mu_5/(2\pi^2\hbar^2v_F c)$.
The coefficients $\sigma^{(B)}$ and $\sigma^{(\epsilon, B)}$ agree with those obtained in
Refs.~\cite{Son:2012wh,Landsteiner:2012kd,Stephanov:2015roa} in the ``no-drag" frame \cite{Rajagopal:2015roa,Stephanov:2015roa,Sadofyev:2015tmb}.

We would like to point out that the Chern--Simons terms $\rho_{\text{{\tiny CS}}}$ and
$\mathbf{j}_{\text{{\tiny CS}}}$ do not directly contribute to the hydrodynamic parts of the charge
and current densities in the Euler equation
(\ref{MHD-Euler}) and the energy conservation relation (\ref{MHD-energy}). Conceptually,
this is the consequence of the specific topological origin of $\rho_{\text{{\tiny CS}}}$ and
$\mathbf{j}_{\text{{\tiny CS}}}$ associated with the filled
electron states deep below the Fermi surface.
In this connection, we note that Eqs.~(\ref{MHD-Euler}) and (\ref{MHD-energy}) originate only from the states near the Fermi surface (see, also, the Supplemental Material).
Indeed, this follows from the fact that the corresponding chiral kinetic equations contain the derivatives from the Fermi--Dirac distribution functions that are insensitive to the details of the energy spectrum at the bottom of the valence band.
Therefore, it is sufficient to use the linearized theory for the matter parts of the consistent hydrodynamics.

One of the key features of the Euler equation (\ref{MHD-Euler}) is the dissipative term proportional
to $\mathbf{u}/\tau$ on its right-hand side, which was also introduced phenomenologically in Ref.~\cite{Landsteiner:2014vua}.
In the hydrodynamic regime, when the electron-electron
scattering is primarily responsible for the formation of the electron fluid, such a term captures
the dominant dissipative effects due to electron scattering on phonons and impurities \cite{Gurzhi}
and explicitly breaks the Galilean invariance. From a physics viewpoint, this is connected with the
existence of the preferred frame in which the ion lattice of a solid is at rest. In the absence of
electromagnetic fields, this term ensures that $\mathbf{u}=\mathbf{0}$ in the global equilibrium state of the
electron fluid.

It should be emphasized that the hydrodynamic equations (\ref{MHD-Euler}) and (\ref{MHD-energy}) lead to the conventional
Ohm's law and Joule's heating. Indeed, in the presence of an external electric field, the
steady state of the electron fluid is reached when the right-hand side of the Euler equation
(\ref{MHD-Euler}) vanishes, i.e., at $\mathbf{u}_{\rm ave} = - e n \tau v_F^2\mathbf{E}/(\epsilon+P)$.
The latter is analogous to the average velocity in the Drude model and reproduces
Ohm's law when the definition for the electric current $\mathbf{J} = -en \mathbf{u}_{\rm ave}$
is taken into account. By substituting $\mathbf{u}_{\rm ave}$ into Eq.~(\ref{MHD-energy}),
one also reproduces the correct local form of Joule's heating.

{\em Maxwell's equations.---}
Since the electron fluid carries a nonzero charge, the hydrodynamic set of equations is
incomplete without Maxwell's equations, i.e.,
\begin{eqnarray}
\label{MHD-Maxwell-be}
\varepsilon_e\bm{\nabla}\cdot\mathbf{E} &=& 4\pi (\rho+\rho_{\rm b}), \\
\label{MHD-Maxwell-bb}
\bm{\nabla}\times\mathbf{E} &=& -\frac{1}{c}\frac{\partial \mathbf{B}}{\partial t}, \\
\frac{1}{\mu_{m}}\bm{\nabla}\times\mathbf{B} &=& \frac{4\pi}{c}\mathbf{J}+\varepsilon_e\frac{1}{c}\frac{\partial \mathbf{E}}{\partial t},
\label{MHD-Maxwell-ee}
\end{eqnarray}
together with $\bm{\nabla}\cdot\mathbf{B} = 0$. Here $\varepsilon_e$ and $\mu_{m}$ denote the electric permittivity and magnetic permeability,
respectively, which originate from the nonitinerant electrons. We note that the
complete electric charge density in Gauss's law (\ref{MHD-Maxwell-be}) must include the contribution of
the electrons in the inner shells and the ions of the lattice $\rho_{\rm b}$.
The latter ensures that the Weyl material is electrically neutral in equilibrium, i.e., $\rho+\rho_{\rm b}=0$.

Maxwell's equations should be supplemented by the expressions for the \emph{total} electron charge and
current densities that include both the hydrodynamic contributions and the Chern--Simons terms, i.e.,
\begin{eqnarray}
\label{MHD-J0-def}
\rho &=& -e n +\frac{\left(\mathbf{B}\cdot\mathbf{u}\right) \sigma^{(B)}}{3v_F^2} + \rho_{\text{{\tiny CS}}},\\
\label{MHD-J-def}
\mathbf{J} &=&-e n\mathbf{u} +\sigma^{(B)} \mathbf{B} +\mathbf{j}_{\text{{\tiny CS}}}.
\end{eqnarray}
In order to be consistent with Eqs.~(\ref{MHD-Maxwell-be}) and (\ref{MHD-Maxwell-ee}),
these electric charge and current densities should satisfy the usual continuity relation
$\partial_t \rho+\bm{\nabla}\cdot \mathbf{J} =0$.
(Note that the Chern--Simons contributions by themselves satisfy the continuity relation.)

Because of the chiral nature of the electron fluid in Weyl materials, the complete set of
hydrodynamic equations should also include the anomalous continuity relation
for the chiral charge,
\begin{equation}
\label{MHD-conserv-eqs-J5}
\partial_t \rho_5 +\bm{\nabla}\cdot \mathbf{J}_5 =-\frac{e^3}{2\pi^2\hbar^2c} \left(\mathbf{E}\cdot\mathbf{B}\right),
\end{equation}
where the chiral charge and current densities are
\begin{eqnarray}
\label{MHD-J50-def}
\rho_5 &=& -en_5 +\frac{\left(\mathbf{B}\cdot\mathbf{u}\right) \sigma^{(B)}_5}{3v_F^2},\\
\label{MHD-J5-def}
\mathbf{J}_5 &=&-e n_5\mathbf{u} +\sigma^{(B)}_5 \mathbf{B}.
\end{eqnarray}
Here $\sigma^{(B)}_5 = e^2\mu/(2\pi^2\hbar^2c)$ is the anomalous transport coefficient
responsible for the chiral separation effect \cite{Vilenkin:1980fu,Zhitnitsky,Newman}.

The complete set of the hydrodynamic and Maxwell's equations presented above is one of the key
results of this paper. Unlike the previous formulations of such equations in the literature,
it incorporates several distinctive features of the chiral electron fluid in Weyl semimetals:
(i) the Chern--Simons contributions affecting Maxwell's equations and
(ii) the broken Galilean invariance due to the ion lattice and the electron scattering on
phonons/impurities.

It is interesting to explore specific observable predictions of the proposed
framework. One of them is the unusual spectrum of collective modes in Weyl semimetals.
A comprehensive study of such modes will be presented elsewhere. Here, in order to
support the general claim and to illuminate the vital role of the Chern--Simons contributions,
we will consider only a few specific modes that propagate transverse to the direction of a static background magnetic field
$\mathbf{B}_0\parallel\hat{\mathbf{z}}$.

{\em Transverse collective excitations.---}
In the state of local equilibrium, the local chemical potentials $\mu$ and $\mu_{5}$ deviate from $\mu_0$ and $\mu_{5,0}$ in global equilibrium. The latter state is
characterized by the vanishing electric current density \cite{Landsteiner:2016,Basar,Franz:2013}
\begin{equation}
\label{ColEx-J0-zero}
\mathbf{J}_0 = \left(\sigma^{(B)}  -\frac{e^3}{2\pi^2\hbar^2c}b_0 \right)\mathbf{B}_0=\mathbf{0},
\end{equation}
and by the condition of electric neutrality,
\begin{equation}
-en_{0}+\rho_{\text{{\tiny CS}}}+\rho_{\rm b}=0.
\label{MHD-mu-def}
\end{equation}
Equation (\ref{ColEx-J0-zero}) is satisfied by setting $\mu_{5,0}=eb_0$. Enforcing Eq.~(\ref{MHD-mu-def})
fixes the value of $\mu_0$, which becomes a function of
temperature, the external magnetic field $\mathbf{B}_0$, the chiral shift $\mathbf{b}$,
and the energy separation $b_0$.

In the study of collective modes, the deviations of the local thermodynamic parameters from
their equilibrium values are small. Then, the use of linearized hydrodynamic equations
is sufficient. They are obtained by looking for a solution in the
form of plane waves, i.e., $\delta\mu(\mathbf{r})=\delta\mu\, e^{-i\omega t+i\mathbf{k}\mathbf{r}}$
together with similar expressions for other oscillating variables (see Sec.~II in the Supplemental Material).
In such an approach, we analyze the spectrum of transverse
(i.e., $\mathbf{k} \perp \mathbf{B}_0$) collective excitations in Weyl semimetals.

By solving the characteristic equation at $n_0=n_{5,0}=0$, we find
the dissipative magnetoacoustic waves with the dispersion relations
given by
\begin{equation}
\label{ColEx-helicon-mumu5=0-s}
\omega_{\rm s,\pm} = -\frac{i}{2\tau} \pm
\frac{i}{2\tau} \sqrt{1 -\frac{4}{3} \tau^2 v_F^2k_{\perp}^2 + \frac{8\tau^2 v_F^2k_{\perp}^2\sigma^{(\epsilon, u)} B_0^2}{3(\epsilon+P)}}.
\end{equation}
Most importantly, there are also collective waves that are strongly affected by the chiral shift $\mathbf{b}$.
One of such modes is realized
when the wave vector $\mathbf{k}$ is parallel to the chiral shift (i.e., $\mathbf{k}\parallel \mathbf{b}\perp
\mathbf{B}_0$ and $b=b_{\perp}$). The corresponding collective excitation has the following
dispersion relation linear in $k_{\perp}$:
\begin{equation}
\label{ColEx-tAHW-mumu5=0-bx}
\omega_{\text{{\tiny tAHW}}}= \frac{c|k_{\perp}|\sqrt{3v_F^3\hbar^3}B_0}{\sqrt{\mu_m} \sqrt{4\pi e^2T^2 b_{\perp}^2 +3\varepsilon_e v_F^3\hbar^3 B_0^2}} +O(k_{\perp}^2).
\end{equation}
As we show below, the propagation of such a mode in Weyl semimetals is sustained by local
currents associated with the anomalous Hall effect (AHE). Therefore, we call it the
\emph{transverse anomalous Hall wave} (tAHW).
(We note that the longitudinal anomalous Hall waves also exist and their study will be reported elsewhere.)

To clarify the physical origin of the tAHW, it is instructive to present the relevant hydrodynamic and Maxwell's
equations. By setting
$\mu_0=\mu_{5,0}=0$
and $\mathbf{b}\perp\mathbf{B}_0$, we arrive at
\begin{eqnarray}
\label{WMHD-B-kx-all-bx-mu0mu50-AHMHDW-system-be}
&&\frac{4\epsilon \omega}{T} \delta T - k_{\perp}\left(\epsilon+P-2B_0^2\sigma^{(\epsilon, u)}\right) \delta u_{\perp} =0,\\
\label{WMHD-B-kx-all-bx-mu0mu50-AHMHDW-system-2}
&&\frac{k_{\perp}}{T} \delta T - \frac{i+\omega\tau}{v_F^2\tau} \delta u_{\perp}
+\frac{5ck_{\perp}^2B_0 \sigma^{(\epsilon, u)}}{\omega (\epsilon+P)} \delta \tilde{E}_{\perp} =0,\\
\label{WMHD-B-kx-all-bx-mu0mu50-AHMHDW-system-3}
&&\frac{T^2 \omega}{3v_F^3\hbar}\delta\mu_5 -i \frac{e^2B_0}{2\pi^2c} \delta E_{\parallel} =0,\\
\label{WMHD-B-kx-all-bx-mu0mu50-AHMHDW-system-4}
&&\left(\omega^2  -\frac{c^2k^2_{\perp}}{\varepsilon_e\mu_m}\right)\delta \tilde{E}_{\perp}
- \frac{2i e^3\omega b_{\perp}}{\pi c\varepsilon_e \hbar^2}\delta E_{\parallel}=0,\\
\label{WMHD-B-kx-all-bx-mu0mu50-AHMHDW-system-ee}
&&\left(\omega^2  -\frac{c^2k^2_{\perp}}{\varepsilon_e\mu_m}\right)\delta E_{\parallel}
+\frac{2i\,e^2\omega }{\pi c\varepsilon_e \hbar^2}\left(B_0\delta \mu_5 +eb_{\perp}\delta \tilde{E}_{\perp}\right) =0,
\nonumber \\
&&
\end{eqnarray}
where subscripts $\parallel$ and $\perp$ denote the vector components parallel and
perpendicular to $\mathbf{B}_0$. In addition, $\delta \tilde{E}_{\perp}$ denotes the component
of the oscillating electric field perpendicular to both $\mathbf{B}_0$ and $\mathbf{b}$. The first two equations originate from the energy conservation relation
and the Euler equation, respectively. The third one is the anomalous continuity relation for the chiral charge and the
last two equations are Maxwell's equations. In the derivation, we also used the Faraday's
law $\delta \mathbf{B}=(c/\omega) [\mathbf{k}\times \delta \mathbf{E}]$ and took
into account that the only oscillating variables in the tAHW are $\delta \mu_5$, $\delta T$, $\delta u_{\perp}$, $\delta E_{\parallel}$, and $\delta \tilde{E}_{\perp}$.

The tAHW is a rather unusual mode that relies on the dynamical electromagnetism
and the Chern--Simons currents in the chiral electron fluid. Its unique nature is clear from
the modified Maxwell's equations (\ref{WMHD-B-kx-all-bx-mu0mu50-AHMHDW-system-4}) and (\ref{WMHD-B-kx-all-bx-mu0mu50-AHMHDW-system-ee}) in which the AHE mixes the transverse and longitudinal components of the oscillating electric field.
(Note that such a mixing occurs even in the absence of an external magnetic field.)
Because of the presence of the magnetic field $\mathbf{B}_0$, the oscillations of the electric fields, in turn, drive the local oscillations of the chiral charge, temperature, and the fluid velocity.
In essence, therefore, the tAHW is a collective excitation that is strongly affected by the topological AHE currents in Weyl semimetals.

\begin{figure}[t]
\begin{center}
\includegraphics[width=\columnwidth]{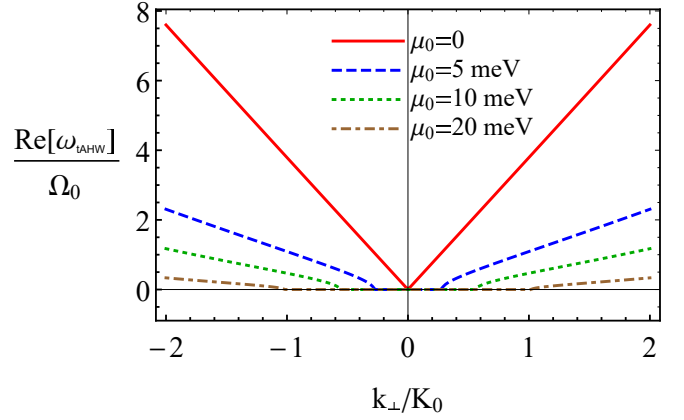}
\caption{(Color online) The real part of the tAHW frequency for several values of $\mu_0$.
The values of the parameters are $T_0=10~\mbox{K}$, $B_0=10^{-2}~\mbox{T}$,
$b=b_{\perp}=0.3\times\pi \hbar c/(e a)$, $\mu_{5,0}=0$, $a = 25.5\times10^{-8}~\mbox{cm}^{-1}$, and $\tau=10^{-12}~\mbox{s}$ \cite{Neupane:2014}. The frequency and the wave vector are
given in units of $\Omega_0=10^{-6}v_F\pi/a\approx 1.84~\mbox{GHz}$ and
$K_{0}=10^{-6}\pi/a\approx 12.3~\mbox{cm}^{-1}$, respectively.}
\label{fig:ColEx-Omega}
\end{center}
\end{figure}

The representative dispersion relation of the tAHW for several values of the chemical potential $\mu_0$ is
shown in Fig.~\ref{fig:ColEx-Omega}. Because
of a finite relaxation time $\tau$, the frequencies generically get nonzero imaginary
parts.
Although this is always the case for the magnetoacoustic waves, the tAHW at the neutrality point $\mu_0=\mu_{5,0}=0$ is not affected by the dissipation effects encoded in $\tau$.
By taking into account that the propagation of the tAHW
is accompanied by oscillations of the electron fluid, this fact is quite amazing.

%%%%%%%%%%%%%%%%%%
%%%%%%%%%%%%%%%%%%

From Fig.~\ref{fig:ColEx-Omega}, we see that the linear dispersion relation of the tAHW is
transformed into a quadratic one at nonzero values of $\mu_0$.
In addition, the frequency of the wave obtains a small but nonzero imaginary part,
i.e., the tAHW becomes dissipative. Let us note that, at small enough values of the wave vector, the real part of
the tAHW vanishes and the mode becomes completely diffusive at $\mu_0\neq0$.

{\em Weyl semimetals with multiple pairs of Weyl nodes.---}
So far, we discussed only the simplest model of a Weyl semimetal with a broken TR
symmetry and a single pair of Weyl nodes. Most Weyl materials, however, have multiple
pairs of Weyl nodes. Moreover, some of them have a broken PI, but intact TR symmetry
(see, e.g., Ref.~\cite{Armitage-Vishwanath:2017-Rev}). When the TR symmetry is present,
the total number of Weyl nodes is a multiple of 4 implying that the sum of all chiral shifts vanishes $\sum_{n} \mathbf{b}^{(n)} =\mathbf{0}$ (here the sum runs over all pairs of Weyl nodes). If this is the case, the net sum of all
Chern--Simons contributions, which are linear in the chiral shifts $\mathbf{b}^{(n)}$, should
vanish too. Then, all distinctive features stemming from a nonzero
$\mathbf{b}$ will also disappear. Of course, this does not apply to Weyl semimetals in which
both the PI and TR symmetries are broken. The hydrodynamic properties of such
materials should be similar to those with only a broken TR symmetry, albeit with the chiral shift
replaced by $\mathbf{b}_{\rm eff}\equiv \sum_{n} \mathbf{b}^{(n)} \neq \mathbf{0}$.

{\em Experimental setup.---}
We would like to note that the existence of the tAHW and other collective modes can be tested experimentally in
Weyl materials with a broken TR symmetry. As in to usual metals \cite{Maxfield}, an experimental
setup requires measuring the transmission amplitude of an electromagnetic wave through a Weyl
crystal as a function of an applied magnetic field or as a function of the frequency at a fixed field.
Because of the interference of standing waves inside the sample, the resulting signal should oscillate
with the magnetic field. The effects of the chiral shift can be studied by changing the
orientation of the crystal and/or magnetic field.

The promising materials for studying the effects of the Chern--Simons terms in the hydrodynamic regime might be magnetic Heusler compounds with a broken TR symmetry \cite{Hasan-magnetic:2016,Cava-Bernevig:2016}. Moreover, such Weyl semimetals have only one pair of well-separated Weyl nodes near the Fermi level. One might also use antiferromagnetic half-Heusler compounds, which are predicted to be Weyl semimetals in an applied magnetic field  \cite{Hirschberger-Ong:2016,Shekhar-Felser:2016,Suzuki:2016}.

{\em Conclusion.---}
In this paper, we formulated a consistent hydrodynamic framework for the chiral electron fluid
in Weyl semimetals. It systematically incorporates the Chern--Simons contributions, the vorticity effects, and takes into
account the breaking of the Galilean invariance by the ion lattice. As argued, the topological
Chern--Simons terms affect the electron motion in the hydrodynamic regime only indirectly via Maxwell's equations.
Nevertheless, they lead to observable effects that are sensitive to the energy and momentum
separations between the Weyl nodes.

By making use of the proposed hydrodynamic theory, we studied the transverse (with respect
to the background magnetic field $\mathbf{B}_0$) collective excitations in Weyl materials. In addition
to the magnetoacoustic waves (which are not modified by the chiral shift $\mathbf{b}$ at the neutrality point), a class
of modes that are significantly altered by the Chern--Simons terms was found.
As a striking example, we mentioned a rather unusual
transverse anomalous Hall wave obtained at $\mathbf{k}\parallel\mathbf{b}$.
At the neutrality point, the latter is a gapless mode with a linear dispersion relation that remains dissipationless even
at finite values of the relaxation time. Most interestingly, the anomalous hall effect currents play the key role in the
physics of the tAHW. At nonzero values of the electric charge density, the frequency of the tAHW becomes quadratic in the wave vector and acquires a small imaginary part.

Our study in this paper was limited to the case of transverse collective excitations.
This was sufficient to illustrate the potential importance of the topological Chern--Simons terms.
Nevertheless, it would be interesting to investigate the spectrum of collective modes in the most general case by using the consistent hydrodynamic theory proposed here.
Such a study is underway and will be reported elsewhere.

\begin{acknowledgments}
The work of E.V.G. was supported partially by the Program of Fundamental Research
of the Physics and Astronomy Division of the NAS of Ukraine.
The work of V.A.M. and P.O.S. was supported by the Natural Sciences and Engineering Research Council of Canada.
The work of I.A.S. was supported by the U.S. National Science Foundation under Grants PHY-1404232
and PHY-1713950.
\end{acknowledgments}

%\end{document}

\begin{widetext}
\newpage
\begin{center}
\textbf{\large \mytitle~--- Supplemental Material} \\[4pt]
E.~V.~Gorbar,$^{1,2}$
V.~A.~Miransky,$^{3}$
I.~A.~Shovkovy,$^{4,5}$
and P.~O.~Sukhachov$^{3}$\\[4pt]
$^{1}${\small\it Department of Physics, Taras Shevchenko National Kiev University, Kiev, 03680, Ukraine}\\
$^{2}${\small\it Bogolyubov Institute for Theoretical Physics, Kiev, 03680, Ukraine}\\
$^{3}${\small\it Department of Applied Mathematics, Western University, London, Ontario, Canada N6A 5B7}\\
$^{4}${\small\it College of Integrative Sciences and Arts, Arizona State University, Mesa, Arizona 85212, USA}\\
$^{5}${\small\it Department of Physics, Arizona State University, Tempe, Arizona 85287, USA}
\end{center}

\setcounter{page}{1}

\input{Supplemental-text-v12}
\end{widetext}

\end{document}

%% file: Supplemental-text-v12.tex
\setcounter{equation}{0}
\setcounter{figure}{0}
\renewcommand{\theequation}{S\arabic{equation}}
\renewcommand{\thefigure}{S\arabic{figure}}

\section{I. Derivation of the hydrodynamic equations}
\label{sec:app-derivation}

Following the standard approach [S1, S2],
we derive the Euler equation and the energy conservation relation (see Secs.~I.1. and I.2., respectively) by calculating the
corresponding moments of the Boltzmann equation of the chiral kinetic theory (CKT) [S3, S4]
in the relaxation time approximation with quasiparticle momentum and energy.
In addition, we provide the explicit expressions for the electric and chiral current densities in the third subsection. The
Boltzmann equation of the CKT in the relaxation time approximation reads
\begin{eqnarray}
&&\left[1-\frac{e}{c}(\mathbf{B}\cdot\mathbf{\Omega}_{\lambda})\right]\frac{\partial f_{\lambda}}{\partial t}+
\left\{-e\tilde{\mathbf{E}}_{\lambda}
-\frac{e}{c}[\mathbf{v}_p\times \mathbf{B}]
+\frac{e^2}{c}(\tilde{\mathbf{E}}_{\lambda} \cdot\mathbf{B})\mathbf{\Omega}_{\lambda}\right\}\cdot
{\bm{\nabla}_{\mathbf{p}} f_{\lambda}} +\left\{\mathbf{v}_p -e[\mathbf{E}\times\mathbf{\Omega}_{\lambda}]
-\frac{e}{c}(\mathbf{v}_p\cdot\mathbf{\Omega}_{\lambda})\mathbf{B}\right\}
\cdot\bm{\nabla}_{\mathbf{r}} f_{\lambda}\nonumber\\
&&=-\frac{f_{\lambda}-f_{\lambda}^{(0)}}{\tau},
\label{app-hydro-der-CKT-Eq}
\end{eqnarray}
where $\tilde{\mathbf{E}}_{\lambda} = \mathbf{E}+(1/e)\partial_\mathbf{r}\epsilon_{p}$, $\mathbf{E}$ and $\mathbf{B}$ are the electric and
magnetic fields, respectively, $\lambda=\pm$ is the chirality of the left-($\lambda=-$) and right-handed ($\lambda=+$) quasiparticles, $-e$ is the electron charge, $\tau$ is
the relaxation time of the intravalley processes. (For the sake of simplicity, we will ignore the chirality-flipping or intervalley processes whose
relaxation time is usually much larger than that of the intravalley ones.)
Since Eq.~(\ref{app-hydro-der-CKT-Eq}) contains the derivatives of the distribution functions that are localized near the Fermi surface, it describes only the ``matter" part of Weyl semimetals properties and is insensitive to the states deep below the Fermi level.
Such a ``matter" response can be effectively described by using the linearized chiral kinetic theory.
While this approach deteriorates far from Weyl nodes, it greatly simplifies the calculations of the low-energy properties.
On the other hand, a two-band model of a Weyl semimetal is crucial to obtain the contribution of the filled electron states deep below the Fermi surface that leads to the topological Chern--Simons terms.
Note that such terms are absent in the linearized chiral kinetic theory.

In order to study the hydrodynamic effects, we seek the distribution function in the local equilibrium form
\begin{equation}
f_{\lambda} = \frac{1}{1+e^{\left(\epsilon_{p}-(\mathbf{u}\cdot\mathbf{p}) -\lambda \hbar \left(\mathbf{p}\cdot\bm{\omega}\right)/(2p)
-\mu_{\lambda}\right)/T}},
\label{app-hydro-der-Fermi-Dirac}
\end{equation}
where $\mathbf{u}$ is the fluid velocity, $\mu_{\lambda}=\mu+\lambda\mu_5$ is the effective chemical potential for the
quasiparticles of a given chirality $\lambda$ containing the electric
$\mu$ and chiral $\mu_5$ chemical potentials, $T$ is temperature, and the term $\lambda \hbar \left(\mathbf{p}\cdot\bm{\omega}\right)/(2p)$ describes the
vortical effects connected with the vorticity $\bm{\omega}=\left[\bm{\nabla}\times\mathbf{u}\right]/2$ of the flow. The quasiparticle energy
reads
\begin{equation}
\epsilon_{p}= v_Fp\left[1 + \frac{e}{c}(\mathbf{B}\cdot \mathbf{\Omega}_{\lambda})\right],
\label{app-hydro-der-epsilon_p}
\end{equation}
where $v_F$ is the Fermi velocity, $c$ is the speed of light,
\begin{equation}
\mathbf{\Omega}_{\lambda} =\lambda \hbar\frac{\hat{\mathbf{p}}}{2p^2}
\label{app-hydro-der-Berry}
\end{equation}
is the Berry curvature [S5], %\cite{Berry:1984}
$p\equiv|\mathbf{p}|$, and $\hat{\mathbf{p}}=\mathbf{p}/p$.
Note that one should make the replacements $\mathbf{\Omega}_{\lambda}\to-\mathbf{\Omega}_{\lambda}$, $\lambda\to-\lambda$, and
$e\to-e$ for the holes (antiparticles). Further, the quasiparticle velocity $\mathbf{v}_p$ equals
\begin{equation}
\mathbf{v}_p= \partial_\mathbf{p}\epsilon_{p}
=v_F\hat{\mathbf{p}} \left[1-2\frac{e}{c} \left(\mathbf{B} \cdot \mathbf{\Omega}_{\lambda}\right) \right]
+ \frac{e v_F}{c}\mathbf{B}\left(\hat{\mathbf{p}} \cdot \mathbf{\Omega}_{\lambda}\right).
\label{app-hydro-der-v}
\end{equation}
For small magnetic field and fluid velocity, we will use the following expansion for the distribution function:
\begin{equation}
f_{\lambda} \approx f_{\lambda}^{(0)} -(\mathbf{p}\cdot\mathbf{u})\frac{\partial f_{\lambda}^{(0)}}{\partial \epsilon_p} +\frac{e}{c}v_Fp(\mathbf{B}\cdot\bm{\Omega}_{\lambda})\frac{\partial f_{\lambda}^{(0)}}{\partial \epsilon_p} -\frac{\lambda \hbar \left(\mathbf{p}\cdot\bm{\omega}\right)}{2p}\frac{\partial f_{\lambda}^{(0)}}{\partial \epsilon_p},
\label{app-hydro-der-f-exp}
\end{equation}
where
\begin{equation}
f_{\lambda}^{(0)} = \frac{1}{1+e^{(v_Fp-\mu_{\lambda})/T}}.
\label{app-hydro-der-f0}
\end{equation}

\subsection{I.1. Derivation of the Euler equation}
\label{sec:app-Euler-der}

In this subsection, we derive the Euler equation by averaging the Boltzmann equation of the CKT (\ref{app-hydro-der-CKT-Eq}) with the quasiparticle momentum $\mathbf{p}$.

\subsubsection{I.1.a. Temporal and momentum derivatives}
\label{sec:app-Euler-der-tmd}

Let us start from the term with the temporal derivative, i.e., the first term in Eq.~(\ref{app-hydro-der-CKT-Eq}). It reads as
\begin{eqnarray}
&&\int\frac{d^3p}{(2\pi\hbar)^3}\mathbf{p}\left[1-\frac{e}{c}(\mathbf{B}\cdot\mathbf{\Omega}_{\lambda})\right]\frac{\partial f_{\lambda}}{\partial t} \approx \int\frac{d^3p}{(2\pi\hbar)^3}\mathbf{p}\left[1-\frac{e}{c}(\mathbf{B}\cdot\mathbf{\Omega}_{\lambda})\right] \nonumber\\
&&\times\frac{\partial }{\partial t} \left[f_{\lambda}^{(0)} -(\mathbf{p}\cdot\mathbf{u})\frac{\partial f_{\lambda}^{(0)}}{\partial \epsilon_p} +\frac{e}{c}v_Fp(\mathbf{B}\cdot\bm{\Omega}_{\lambda})\frac{\partial f_{\lambda}^{(0)}}{\partial \epsilon_p} -\frac{\lambda \hbar \left(\mathbf{p}\cdot\bm{\omega}\right)}{2p}\frac{\partial f_{\lambda}^{(0)}}{\partial \epsilon_p}\right].
\label{app-Euler-der-1}
\end{eqnarray}
Here, we have
\begin{eqnarray}
\label{app-Euler-der-1-2}
&&-\int\frac{d^3p}{(2\pi\hbar)^3}\mathbf{p}(\mathbf{p}\cdot\mathbf{u})\frac{\partial f_{\lambda}^{(0)}}{\partial \epsilon_p}=-\frac{\mathbf{u}}{3} \int \frac{d^3p\,p^2}{(2\pi\hbar)^3} \frac{\partial f_{\lambda}^{(0)}}{\partial \epsilon_p} = -\frac{\mathbf{u}}{3} \frac{T^4 4!}{2\pi^2 v_F^5 \hbar^3} \mathrm{Li}_4\left(-e^{\mu_{\lambda}/T}\right) \stackrel{\sum_{p,h} \sum_{\lambda}}{=}  \frac{\mathbf{u} (\epsilon+P)}{v_F^2},\\
\label{app-Euler-der-1-3}
&&v_F\frac{e}{c}\int\frac{d^3p}{(2\pi\hbar)^3}\mathbf{p}p(\mathbf{B}\cdot\bm{\Omega}_{\lambda})\frac{\partial f_{\lambda}^{(0)}}{\partial \epsilon_p}
= \frac{\mathbf{B} \hbar v_F e \lambda}{6c}\frac{T^2 2!}{2\pi^2 v_F^3 \hbar^3} \mathrm{Li}_2\left(-e^{\mu_{\lambda}/T}\right)
\stackrel{\sum_{p,h} \sum_{\lambda}}{=} -\frac{e\mu\mu_5 \mathbf{B}}{3c\pi^2v_F^2 \hbar^2},\\
\label{app-Euler-der-1-3-vort}
&&-\frac{\lambda \hbar}{2}\int\frac{d^3p}{(2\pi\hbar)^3}\mathbf{p}\frac{\left(\mathbf{p}\cdot\bm{\omega}\right)}{p}\frac{\partial f_{\lambda}^{(0)}}{\partial \epsilon_p}
= -\frac{\lambda \hbar \bm{\omega}}{6} \frac{T^3 3!}{2\pi^2 v_F^4 \hbar^3} \mathrm{Li}_3\left(-e^{\mu_{\lambda}/T}\right) \stackrel{\sum_{p,h} \sum_{\lambda}}{=} \frac{\hbar \bm{\omega} n_{5}}{2v_F},\\
\label{app-Euler-der-1-4}
&&-\frac{e}{c}\int\frac{d^3p}{(2\pi\hbar)^3}(\mathbf{B}\cdot\bm{\Omega}_{\lambda})\mathbf{p}f_{\lambda}^{(0)} =-\frac{\mathbf{B} \hbar e \lambda}{6c} \int \frac{d^3p}{(2\pi\hbar)^3} \frac{f_{\lambda}^{(0)}}{p}
= \frac{\mathbf{B} \hbar e \lambda}{6c} \frac{T^2}{2\pi^2 v_F^2} \mathrm{Li}_2\left(-e^{\mu_{\lambda}/T}\right) \stackrel{\sum_{p,h} \sum_{\lambda}}{=} \frac{e\mu\mu_5 \mathbf{B}}{6c\pi^2v_F^2  \hbar^2},
\end{eqnarray}
where $\sum_{p,h} $ denotes the summation over particles and holes, $\mbox{Li}_n(x)$ is the
polylogarithm function, and we used the formulas from Sec.~III. In addition,
\begin{eqnarray}
\label{app-Euler-der-equilibrium-be}
\epsilon &\simeq& \sum_{\lambda=\pm} \frac{15\mu_{\lambda}^4+30\pi^2T^2\mu_{\lambda}^2+7\pi^4T^4}{120\pi^2\hbar^3v_F^3}= \frac{\mu^4+6\mu^2\mu_{5}^2+\mu_{5}^4}{4\pi^2\hbar^3v_F^3} +\frac{T^2(\mu^2+\mu_{5}^2)}{2\hbar^3v_F^3} +\frac{7\pi^2T^4}{60\hbar^3v_F^3},\\
P &\simeq& \frac{\epsilon}{3},\\
n &\simeq& \frac{1}{\hbar^3v_F^3}\sum_{\lambda=\pm} \mu_{\lambda}\left(\frac{\mu_{\lambda}^2}{6\pi^2} +\frac{T^2}{6}\right) = \frac{\mu\left(\mu^2+3\mu^2_{5}+\pi^2T^2\right)}{3\pi^2 \hbar^3 v_F^3 },\\
n_5 &\simeq& \frac{1}{\hbar^3 v_F^3}\sum_{\lambda=\pm} \lambda\mu_{\lambda}\left(\frac{\mu_{\lambda}^2}{6\pi^2} +\frac{T^2}{6}\right) = \frac{\mu_{5}\left(\mu^2_{5}+3\mu^2+\pi^2T^2\right)}{3\pi^2 \hbar^3v_F^3}.
\label{app-Euler-der-equilibrium-ee}
\end{eqnarray}
Here we ignored the corrections of order $B^2$ in the expressions for the energy density, pressure, and number densities.

Thus, the first term in Eq.~(\ref{app-hydro-der-CKT-Eq}) leads to the following contribution of the Euler equation:
\begin{equation}
\partial_t \left[\frac{\mathbf{u} (\epsilon+P)}{v_F^2} -\frac{e\mu\mu_5 \mathbf{B}}{2c\pi^2v_F^2 \hbar^2}
+\frac{\hbar \bm{\omega} n_{5}}{2v_F}\right].
\label{app-Euler-der-1-all}
\end{equation}

Let us consider now the second term in the kinetic equation, which is related to forces acting on quasiparticles, i.e.,
\begin{equation}
\int\frac{d^3p}{(2\pi\hbar)^3}\mathbf{p}\left\{-e\mathbf{E}
-\frac{e}{c}[\mathbf{v}_p\times \mathbf{B}]
+\frac{e^2}{c}(\mathbf{E}\cdot\mathbf{B})\mathbf{\Omega}_{\lambda}\right\}\cdot
{\bm{\nabla}_{\mathbf{p}} f_{\lambda}},
\label{app-Euler-der-2}
\end{equation}
where we have dropped the gradient terms in $\tilde{\mathbf{E}}$, i.e., replaced $\tilde{\mathbf{E}}\to\mathbf{E}$. The corresponding
contribution will be accounted later. Integrating by parts, we find the following ``force" contribution on the left-hand side of the Euler equation:
\begin{equation}
en\mathbf{E} -\frac{e}{c}\left[\mathbf{B}\times \left(n\mathbf{u}
+\frac{\bm{\omega}\mu\mu_5}{3\pi^2v_F^2\hbar^2}
\right)\right] -\frac{e^2\mu_5 \mathbf{u} (\mathbf{E}\cdot\mathbf{B})}{6c\pi^2\hbar^2v_F^2} -\frac{e^2 (\mathbf{E}\cdot\mathbf{B})\bm{\omega}}{24\pi^2c v_F \hbar}.
\label{app-Euler-der-3-all}
\end{equation}
One can easily see that the chiral vortical effect (CVE) current [S6] is also present in Eq.~(\ref{app-Euler-der-3-all}), albeit with the additional multiplier $1/3$. Note also that we omitted the singular terms connected with the divergent integrals in the infrared region. Such terms cannot be reliably
described in the CKT, therefore, we do not consider them.

\subsubsection{I.1.b. Spatial derivatives}
\label{sec:app-Euler-der-sd}

Further, we consider the terms with spatial derivatives in the Euler equation. Let us start from the last term on the left-hand side of the CKT equation (\ref{app-hydro-der-CKT-Eq}), i.e.,
\begin{equation}
\left\{\mathbf{v}_p-e[\mathbf{E}\times\mathbf{\Omega}_{\lambda}]
-\frac{e}{c}(\mathbf{v}_p\cdot\mathbf{\Omega}_{\lambda})\mathbf{B}\right\}
\cdot\bm{\nabla}_{\mathbf{r}} f_{\lambda}.
\label{app-Euler-der-gradient-0}
\end{equation}
Multiplying the above equation
by $\mathbf{p}$ and integrating over the momentum, we obtain
\begin{eqnarray}
\label{app-Euler-der-gradient-all-1}
&&\bm{\nabla}_{\mathbf{r}}P
- \frac{2e}{15\pi^2cv_F^2\hbar^2}  \left[\sum_{j=x,y,z}\mathbf{B}_j \bm{\nabla}_{\mathbf{r}}\mathbf{u}_j +(\mathbf{B}\cdot\bm{\nabla}_{\mathbf{r}})\mathbf{u}
+\mathbf{B} (\bm{\nabla}_{\mathbf{r}}\cdot\mathbf{u})\right] \mu\mu_5
+\frac{e^2v_F}{60\pi^2c^2\hbar} \left[\frac{1}{2}\bm{\nabla}_{\mathbf{r}}B^2 +(\mathbf{B}\cdot\bm{\nabla}_{\mathbf{r}})\mathbf{B} \right]\nonumber\\
&&-\frac{e}{30\pi^2cv_F\hbar} \left[(\mathbf{B}\cdot\bm{\nabla}_{\mathbf{r}})\bm{\omega} + \sum_{j=x,y,z} \mathbf{B}_j\bm{\nabla}_{\mathbf{r}} \bm{\omega}_j  \right] \mu - \frac{e\left[\bm{\nabla}_{\mathbf{r}}\times\mathbf{E}\right] \mu\mu_5}{6\pi^2 v_F^2\hbar^2}.
\end{eqnarray}
Note that the position of the variables is important, because the derivative acts on all variables which are placed next to it.

Last but not least, we study the gradient terms arising from the $1/e \bm{\nabla}_{\mathbf{r}} \epsilon_p$ term in $\tilde{\mathbf{E}}$, i.e.,
\begin{equation}
\label{app-Euler-der-gradient-4-0}
\frac{1}{e} \bm{\nabla}_{\mathbf{r}} \epsilon_p =\sum_{j=x,y,z}\frac{v_Fp}{c}\bm{\Omega}_{\lambda, j} (\bm{\nabla}_{\mathbf{r}}\mathbf{B}_j) = \sum_{j=x,y,z}\frac{v_F \lambda \hbar }{2cp^2} \mathbf{p}_j (\bm{\nabla}_{\mathbf{r}}\mathbf{B}_j).
\end{equation}

By using Eq.~(\ref{app-Euler-der-2}) with $\mathbf{E}\to 1/e \bm{\nabla}_{\mathbf{r}} \epsilon_p$ and integrating over the momentum, we
obtain
\begin{eqnarray}
\label{app-Euler-der-gradient-all-2}
&&-\frac{e^2v_F}{60\pi^2c^2\hbar} \left[\frac{1}{2}(\bm{\nabla}_{\mathbf{r}}B^2) +(\mathbf{B}\cdot\bm{\nabla}_{\mathbf{r}})\mathbf{B}\right] - \sum_{j=x,y,z} \frac{e\mu\mu_5\mathbf{u}_j\bm{\nabla}_{\mathbf{r}}\mathbf{B}_j}{3\pi^2cv_F^2\hbar^2} + \frac{2e\mu\mu_5}{15\pi^2cv_F^2\hbar^2} \left[\sum_{j=x,y,z} \mathbf{u}_j\bm{\nabla}_{\mathbf{r}}\mathbf{B}_j +(\mathbf{u}\cdot\bm{\nabla}_{\mathbf{r}})\mathbf{B}\right] \nonumber\\
&&+ \frac{e^2v_F \bm{\nabla}_{\mathbf{r}}B^2}{48\pi^2c^2\hbar} +\sum_{j=x,y,z} \frac{e\mu\bm{\omega}_j \bm{\nabla}_{\mathbf{r}}\mathbf{B}_j}{12\pi^2cv_F\hbar}
+\frac{e}{30 \pi^2 v_Fc \hbar} \left[\sum_{j=x,y,z}\bm{\omega}_j\bm{\nabla}_{\mathbf{r}}\mathbf{B}_j + (\bm{\omega}\cdot\bm{\nabla}_{\mathbf{r}})\mathbf{B} \right].
\end{eqnarray}

Thus, we find the following gradient terms:
\begin{eqnarray}
\label{app-Euler-der-gradient-all}
&&\bm{\nabla}_{\mathbf{r}}P
- \frac{2e}{15\pi^2cv_F^2\hbar^2}  \left[\sum_{j=x,y,z}\mathbf{B}_j \bm{\nabla}_{\mathbf{r}}\mathbf{u}_j +(\mathbf{B}\cdot\bm{\nabla}_{\mathbf{r}})\mathbf{u}
 +\mathbf{B} (\bm{\nabla}_{\mathbf{r}}\cdot\mathbf{u})\right] \mu\mu_5
- \frac{e\left[\bm{\nabla}_{\mathbf{r}}\times\mathbf{E}\right] \mu\mu_5}{6\pi^2 v_F^2\hbar^2}
- \sum_{j=x,y,z}\frac{e\mu\mu_5\mathbf{u}_j\bm{\nabla}_{\mathbf{r}}\mathbf{B}_j}{3\pi^2cv_F^2\hbar^2} \nonumber\\
&&+ \frac{2e\mu\mu_5}{15\pi^2cv_F^2\hbar^2} \left[\sum_{j=x,y,z}\mathbf{u}_j\bm{\nabla}_{\mathbf{r}}\mathbf{B}_j +(\mathbf{u}\cdot\bm{\nabla}_{\mathbf{r}})\mathbf{B}\right] + \frac{e^2v_F (\bm{\nabla}_{\mathbf{r}}B^2)}{48\pi^2c^2\hbar}
-\frac{e}{30\pi^2cv_F\hbar} \left[(\mathbf{B}\cdot\bm{\nabla}_{\mathbf{r}})\bm{\omega} + \sum_{j=x,y,z} \mathbf{B}_j\bm{\nabla}_{\mathbf{r}} \bm{\omega}_j \right] \mu  \nonumber\\
&&+\sum_{j=x,y,z} \frac{e\mu\bm{\omega}_j \bm{\nabla}_{\mathbf{r}}\mathbf{B}_j}{12\pi^2cv_F\hbar}
+\frac{e}{30 \pi^2 v_Fc \hbar} \left[\sum_{j=x,y,z}\bm{\omega}_j\bm{\nabla}_{\mathbf{r}}\mathbf{B}_j + (\bm{\omega}\cdot\bm{\nabla}_{\mathbf{r}})\mathbf{B} \right].
\end{eqnarray}

\subsubsection{I.1.c. Collision term and the complete Euler equation}
\label{sec:app-Euler-der-collison}

Finally, we should consider the collision integral on the right-hand side of the CKT equation (\ref{app-hydro-der-CKT-Eq}), i.e.,
\begin{eqnarray}
\label{app-Euler-der-collision}
&&-\frac{1}{\tau} \int\frac{d^3p}{(2\pi\hbar)^3} \mathbf{p} \left[1-\frac{e}{c}\left(\mathbf{B}\cdot\bm{\Omega}_{\lambda}\right)\right] \left(f_{\lambda} -f_{\lambda}^{(\rm eq)}\right)
= \frac{1}{\tau} \int\frac{d^3p}{(2\pi\hbar)^3} \mathbf{p} \left[(\mathbf{p}\cdot\mathbf{u})\frac{\partial f_{\lambda}^{(0)}}{\partial \epsilon_p}
+\frac{\lambda \hbar \left(\mathbf{p}\cdot\bm{\omega}\right)}{2p}\frac{\partial f_{\lambda}^{(0)}}{\partial \epsilon_p}\right] \nonumber\\
&&\stackrel{\sum_{p,h} \sum_{\lambda}}{=} -\frac{(\epsilon+P) \mathbf{u}}{v_F^2\tau} - \frac{\hbar \bm{\omega} n_5}{2v_F \tau},
\end{eqnarray}
where $f_{\lambda}^{(\rm eq)}$ is the equilibrium distribution function given in Eq.~(\ref{app-hydro-der-Fermi-Dirac}) at
$\mathbf{u}=\mathbf{0}$.

Finally, the Euler equation derived from the Boltzmann equation of the CKT in the relaxation time approximation is
\begin{eqnarray}
\label{app-Euler-der-result}
&&\frac{1}{v_F}\partial_t \left[\frac{\mathbf{u} (\epsilon+P)}{v_F} -\frac{e\mu\mu_5}{2\pi^2\hbar^2v_F c}\mathbf{B} +\frac{\hbar \bm{\omega} n_{5}}{2}\right] +\bm{\nabla}_{\mathbf{r}}P - \frac{2e}{15\pi^2cv_F^2\hbar^2}  \left[\sum_{j=x,y,z} \mathbf{B}_j \bm{\nabla}_{\mathbf{r}}\mathbf{u}_j + (\mathbf{B}\cdot\bm{\nabla}_{\mathbf{r}})\mathbf{u} +\mathbf{B} (\bm{\nabla}_{\mathbf{r}}\cdot \mathbf{u})\right] \mu\mu_5 \nonumber\\
&&- \frac{e\left[\bm{\nabla}_{\mathbf{r}}\times\mathbf{E}\right] \mu\mu_5}{6\pi^2 v_F^2\hbar^2}
- \sum_{j=x,y,z} \frac{e\mu\mu_5\mathbf{u}_j\bm{\nabla}_{\mathbf{r}}\mathbf{B}_j}{3\pi^2cv_F^2\hbar^2}
+ \frac{2e\mu\mu_5}{15\pi^2cv_F^2\hbar^2} \left[\sum_{j=x,y,z} \mathbf{u}_j\bm{\nabla}_{\mathbf{r}}\mathbf{B}_j +(\mathbf{u}\cdot\bm{\nabla}_{\mathbf{r}})\mathbf{B}\right] + \frac{e^2v_F \bm{\nabla}_{\mathbf{r}}B^2}{48\pi^2c^2\hbar} \nonumber\\
&&-\frac{e}{30\pi^2cv_F\hbar} \left[(\mathbf{B}\cdot\bm{\nabla}_{\mathbf{r}})\bm{\omega} + \sum_{j=x,y,z} \mathbf{B}_j\bm{\nabla}_{\mathbf{r}} \bm{\omega}_j \right] \mu  +\sum_{j=x,y,z} \frac{e\mu\bm{\omega}_j \bm{\nabla}_{\mathbf{r}}\mathbf{B}_j}{12\pi^2cv_F\hbar}
+\frac{e}{30 \pi^2 v_Fc \hbar} \left[\sum_{j=x,y,z} \bm{\omega}_j\bm{\nabla}_{\mathbf{r}}\mathbf{B}_j + (\bm{\omega}\cdot\bm{\nabla}_{\mathbf{r}})\mathbf{B} \right] \nonumber\\
&& =-en\mathbf{E} +\frac{e}{c}\left[\mathbf{B}\times \left\{n\mathbf{u}
+\frac{\bm{\omega}\mu\mu_5}{3\pi^2v_F^2\hbar^2}
\right\}\right] +\frac{e^2\mu_5 \mathbf{u} (\mathbf{E}\cdot\mathbf{B})}{6c\pi^2\hbar^2v_F^2} +\frac{e^2 (\mathbf{E}\cdot\mathbf{B})\bm{\omega}}{24\pi^2c v_F \hbar} -\frac{(\epsilon+P) \mathbf{u}}{v_F^2\tau} - \frac{\hbar \bm{\omega} n_5}{2 v_F \tau},
\end{eqnarray}
where derivatives act on all terms situated to the right of them. The abbreviated version of this equation appears in the main text as Eq.~(5).

\subsection{I.2. Derivation of the energy conservation relation}
\label{sec:app-energy-der}

In this subsection, we derive the energy conservation relation by averaging the CKT equation (\ref{app-hydro-der-CKT-Eq}) with
the quasiparticle energy $\epsilon_p$.

\subsubsection{I.2.a. Temporal and momentum derivatives}
\label{sec:app-energy-der-tmd}

The term with the temporal derivative reads
\begin{eqnarray}
&&\int\frac{d^3p}{(2\pi\hbar)^3}\epsilon_p\left[1-\frac{e}{c}(\mathbf{B}\cdot\mathbf{\Omega}_{\lambda})\right] \partial_t f_{\lambda} \approx \int\frac{d^3p}{(2\pi\hbar)^3}v_Fp\left[1-\frac{e^2}{c^2}(\mathbf{B}\cdot\mathbf{\Omega}_{\lambda})^2\right] \nonumber\\
&&\times \partial_t \left[f_{\lambda}^{(0)} -(\mathbf{p}\cdot\mathbf{u})\frac{\partial f_{\lambda}^{(0)}}{\partial \epsilon_p} +\frac{e}{c}v_Fp(\mathbf{B}\cdot\bm{\Omega}_{\lambda})\frac{\partial f_{\lambda}^{(0)}}{\partial \epsilon_p} -\frac{\lambda \hbar \left(\mathbf{p}\cdot\bm{\omega}\right)}{2p}\frac{\partial f_{\lambda}^{(0)}}{\partial \epsilon_p}\right].
\label{app-energy-der-t}
\end{eqnarray}
Omitting singular (i.e., high order in the Berry curvature) and odd in $\mathbf{p}$ terms, we find the only nontrivial contribution
\begin{equation}
\int\frac{d^3p}{(2\pi\hbar)^3}v_Fp \,\partial_t f_{\lambda}^{(0)} =  -\partial_{t}\frac{T^4 3!}{2\pi^2 v_F^3 \hbar^3} \mathrm{Li}_4\left(-e^{\mu_{\lambda}/T}\right) \stackrel{\sum_{p,h} \sum_{\lambda}}{=} \partial_{t}\epsilon.
\label{app-energy-der-t-1}
\end{equation}

Replacing $\tilde{\mathbf{E}}\to\mathbf{E}$ in the second term with momentum derivatives in the CKT equation
(\ref{app-hydro-der-CKT-Eq}), multiplying it by $\epsilon_p$, and integrating over the momentum, we find the following ``force" term on
the left-hand side of the energy conservation relation:
\begin{equation}
\label{app-energy-der-energy-force}
\mathbf{E}\cdot \left(en\mathbf{u}-\frac{e^2\mu_5}{2\pi^2\hbar^2c}\mathbf{B} +\frac{e \bm{\omega} \mu\mu_5}{3\pi^2v_F^2\hbar^2} \right).
\end{equation}
Here the second term in the parentheses appears due to the chiral magnetic effect (CME) current [S7--S9] and the third one corresponds to the CVE contribution.

\subsubsection{I.2.b. Spatial derivatives}
\label{sec:app-energy-der-sd}

As to the terms with spatial derivatives, we begin with the third term on the left-hand side of Eq.~(\ref{app-hydro-der-CKT-Eq}), i.e.,
\begin{equation}
\left\{\mathbf{v}_p -e[\mathbf{E}\times\mathbf{\Omega}_{\lambda}]
-\frac{e}{c}(\mathbf{v}_p\cdot\mathbf{\Omega}_{\lambda})\mathbf{B}\right\}
\cdot\bm{\nabla}_{\mathbf{r}} f_{\lambda}.
\label{app-energy-der-gradient-0}
\end{equation}

Multiplying it by $\epsilon_p$ and integrating over the momentum, we obtain
\begin{eqnarray}
\label{app-energy-der-result-1}
&&(\bm{\nabla}_{\mathbf{r}}\cdot\mathbf{u}) (\epsilon+P)
+\frac{e^2 v_F}{60\pi^2c^2\hbar} \left[\frac{1}{2}\sum_{j=x,y,z} \mathbf{B}_j \left(\mathbf{B}\cdot\bm{\nabla}_{\mathbf{r}}\right)\mathbf{u}_j -B^2(\bm{\nabla}_{\mathbf{r}}\cdot\mathbf{u})\right]
+ \frac{e \left(\mathbf{E}\cdot\left[\bm{\nabla}_{\mathbf{r}}\times\mathbf{u}\right]\right)\mu\mu_5}{3\pi^2v_F^2\hbar^2}
\nonumber\\
&&-\frac{e^2 v_F \left(\mathbf{E}\cdot\left[\bm{\nabla}_{\mathbf{r}}\times\mathbf{B}\right]\right)}{24\pi^2 \hbar c}
-\frac{e\left(\mathbf{B}\cdot\bm{\nabla}_{\mathbf{r}}\right)\mu\mu_5}{2\pi^2c\hbar^2}
+\frac{\hbar v_F (\bm{\nabla}_{\mathbf{r}}\cdot\bm{\omega})n_5}{2}
- \frac{e\left(\mathbf{E}\cdot\left[\bm{\nabla}_{\mathbf{r}} \times \bm{\omega}\right]\right)\mu}{12\pi^2 v_F \hbar}.
\end{eqnarray}

Finally, we should include the terms with spatial derivatives in $\tilde{\mathbf{E}}$ that gives
\begin{equation}
\label{app-energy-der-result-2}
-\frac{e^2v_F}{60\pi^2c^2\hbar} \left[\left(\mathbf{u}\cdot\bm{\nabla}_{\mathbf{r}}\right)B^2 -3\sum_{j=x,y,z} \mathbf{u}_j\left(\mathbf{B}\cdot\bm{\nabla}_{\mathbf{r}}\right)\mathbf{B}_j\right].
\end{equation}

\subsubsection{I.2.c. Collision term and the complete energy conservation relation}
\label{sec:app-energy-der-collison}

As is expected for elastic collisions, the collision integral in the relaxation time approximation does not contribute to the energy conservation
relation.
Thus, the complete energy conservation equation reads
\begin{eqnarray}
\label{app-energy-der-result}
&&\partial_t \epsilon
+(\bm{\nabla}_{\mathbf{r}}\cdot\mathbf{u}) (\epsilon+P)
+\frac{e^2 v_F}{60\pi^2c^2\hbar} \left[\frac{1}{2}\sum_{j=x,y,z}\mathbf{B}_j \left(\mathbf{B}\cdot\bm{\nabla}_{\mathbf{r}}\right)\mathbf{u}_j -B^2(\bm{\nabla}_{\mathbf{r}}\cdot\mathbf{u})\right] + \frac{e \left(\mathbf{E}\cdot\left[\bm{\nabla}_{\mathbf{r}}\times\mathbf{u}\right]\right)\mu\mu_5}{3\pi^2v_F^2\hbar^2}\nonumber\\
&&-\frac{e^2 v_F \left(\mathbf{E}\cdot\left[\bm{\nabla}_{\mathbf{r}}\times\mathbf{B}\right]\right)}{24\pi^2 \hbar c} -\frac{e\left(\mathbf{B}\cdot\bm{\nabla}_{\mathbf{r}}\right)\mu\mu_5}{2\pi^2c\hbar^2}
-\frac{e^2v_F}{60\pi^2c^2\hbar} \left[\left(\mathbf{u}\cdot\bm{\nabla}_{\mathbf{r}}\right)B^2 -3\sum_{j=x,y,z} \mathbf{u}_j\left(\mathbf{B}\cdot\bm{\nabla}_{\mathbf{r}}\right)\mathbf{B}_j\right]
\nonumber\\
&&+\frac{\hbar v_F (\bm{\nabla}_{\mathbf{r}}\cdot\bm{\omega})n_5}{2} - \frac{e\left(\mathbf{E}\cdot\left[\bm{\nabla}_{\mathbf{r}} \times \bm{\omega}\right]\right)\mu}{12\pi^2 v_F \hbar} =\left(\mathbf{E}\cdot \left[-en\mathbf{u}+\frac{e^2\mu_5}{2\pi^2\hbar^2c}\mathbf{B} - \frac{e \bm{\omega} \mu\mu_5}{3\pi^2v_F^2\hbar^2} \right]\right).
\end{eqnarray}
The abbreviated version of the above relation appears in the main text as Eq.~(6).

\subsection{I.3. Derivation of the electric and chiral current densities}
\label{sec:app-current-der}

In order to finish our derivation of the consistent hydrodynamic equations for the chiral electrons in Weyl materials, we should find the charge and current densities in the presence of the fluid velocity $\mathbf{u}$ and the vorticity $\bm{\omega}$. The topological Chern--Simons parts of the electric charge and current densities (also known as the Bardeen--Zumino terms in high energy physics [S10]) were derived using the quantum field theory approach in Refs.~[S11--S12] as well as in a two band model of a Weyl semimetal in Ref.~[S13] (see also the main text).

In components, the Chern--Simons terms take the following form:
\begin{eqnarray}
\rho_{\text{{\tiny CS}}} &=&- \frac{e^3}{2\pi^2\hbar^2c^2}\,(\mathbf{b}\cdot\mathbf{B}),
\label{app-current-CS-charge}
 \\
\mathbf{J}_{\text{{\tiny CS}}} &=&-\frac{e^3}{2\pi^2\hbar^2c}\,b_0 \mathbf{B} + \frac{e^3}{2\pi^2\hbar^2c}\,\left[\mathbf{b}\times\mathbf{E}\right],
\label{app-current-CS-current}
\end{eqnarray}
where $b_0$ and $\mathbf{b}$ correspond to the energy and momentum separations
between the Weyl nodes.

The matter parts of the electric charge and current densities equal
\begin{equation}
\label{app-current-J0-def}
\rho_{\rm mat} \approx
-e \int\frac{d^3p}{(2\pi\hbar)^3} \left[1-\frac{e}{c}\left(\mathbf{B}\cdot\bm{\Omega}_{\lambda}\right)\right] \left[f_{\lambda}^{(0)} -(\mathbf{p}\cdot\mathbf{u})\frac{\partial f_{\lambda}^{(0)}}{\partial \epsilon_p} +\frac{e}{c}v_Fp(\mathbf{B}\cdot\bm{\Omega}_{\lambda})\frac{\partial f_{\lambda}^{(0)}}{\partial \epsilon_p} -\frac{\lambda \hbar \left(\mathbf{p}\cdot\bm{\omega}\right)}{2p}\frac{\partial f_{\lambda}^{(0)}}{\partial \epsilon_p}\right]
\end{equation}
and
\begin{eqnarray}
\label{app-current-J-def}
\mathbf{J}_{\rm mat} &\approx& -e\int\frac{d^3p}{(2\pi\hbar)^3} \left\{\mathbf{v}_p-e[\mathbf{E}\times\mathbf{\Omega}_{\lambda}]
-\frac{e}{c}(\mathbf{v}_p\cdot\mathbf{\Omega}_{\lambda})\mathbf{B}
\right\} \nonumber\\
&\times&
\left[f_{\lambda}^{(0)} -(\mathbf{p}\cdot\mathbf{u})\frac{\partial f_{\lambda}^{(0)}}{\partial \epsilon_p} +\frac{e}{c}v_Fp(\mathbf{B}\cdot\bm{\Omega}_{\lambda})\frac{\partial f_{\lambda}^{(0)}}{\partial \epsilon_p} -\frac{\lambda \hbar \left(\mathbf{p}\cdot\bm{\omega}\right)}{2p}\frac{\partial f_{\lambda}^{(0)}}{\partial \epsilon_p}\right],
\end{eqnarray}
respectively. Integrating over the momentum, we find
\begin{eqnarray}
\label{app-current-J0-matter-0}
\rho_{\rm mat} &=& -en -\frac{e^2 \mu_5\left(\mathbf{B}\cdot\mathbf{u}\right)}{6\pi^2v_F^2c\hbar^2} + \frac{e^2 \left(\mathbf{B}\cdot\bm{\omega}\right)}{24\pi^2 c v_F \hbar}
-\frac{e\mu \bm{\omega}^2}{6\pi^2v_F^3\hbar},\\
\label{app-current-J-matter-0}
\mathbf{J}_{\rm mat} &=& -en\mathbf{u} -\frac{e\mu\mu_5\bm{\omega}}{3\pi^2v_F^2\hbar^2} +\frac{e^2\mu_5 \mathbf{B}}{2\pi^2\hbar^2c}
+\frac{e^2\mu_5 \left[\mathbf{E}\times\mathbf{u}\right]}{6\pi^2\hbar^2v_F^2}
+ \frac{e^2\left[\mathbf{E}\times\bm{\omega}\right]}{24\pi^2v_F\hbar}.
\end{eqnarray}
Here, we obtained only $1/3$ of the CVE. In order to reproduce it correctly, one should include the magnetization current [S6]
%\cite{Chen:2014cla}
\begin{eqnarray}
\mathbf{J}_{\rm mag}&=&-e\bm{\nabla}\times \int\frac{d^3p}{(2\pi \hbar)^3} f_{\lambda}\epsilon_{\mathbf{p}}\mathbf{\Omega}_{\lambda}
=-e\bm{\nabla}\times \int\frac{d^3p}{(2\pi \hbar)^3} \left[f_{\lambda}^{(0)} -(\mathbf{p}\cdot\mathbf{u})\frac{\partial f_{\lambda}^{(0)}}{\partial \epsilon_p} +\frac{e}{c}v_Fp(\mathbf{B}\cdot\bm{\Omega}_{\lambda})\mathbf{\Omega}_{\lambda}\frac{\partial f_{\lambda}^{(0)}}{\partial \epsilon_p} -\frac{\lambda \hbar \left(\mathbf{p}\cdot\bm{\omega}\right)}{2p}\frac{\partial f_{\lambda}^{(0)}}{\partial \epsilon_p}\right]  \nonumber\\
&\times&\mathbf{\Omega}_{\lambda} v_Fp\left[1 + \frac{e}{c}(\mathbf{B}\cdot \mathbf{\Omega}_{\lambda})\right] = -\frac{e\left[\bm{\nabla}\times\mathbf{u}\right]\mu\mu_5}{3\pi^2 v_F^2\hbar^2} -\frac{e \left[\bm{\nabla}\times\bm{\omega}\right] \mu}{12\pi^2 \hbar v_F}= -\frac{2 e\mu\mu_5\bm{\omega}}{3\pi^2v_F^2\hbar^2} +\frac{e\left[\mathbf{u}\times\bm{\nabla}\right]\mu\mu_5}{3\pi^2 v_F^2\hbar^2} -\frac{e \left[\bm{\nabla}\times\bm{\omega}\right] \mu}{12\pi^2 \hbar v_F},
\label{CKT-electric-current-magnetization-def}
\end{eqnarray}
where the first term in the last equality clearly provides the missing $2/3$ of the CVE.

Thus, the final expressions for the nontopological parts of the electric charge and current densities are
\begin{eqnarray}
\label{app-current-J0-matter}
\rho_{\rm mat} &=& -en -\frac{e^2\mu_5\left(\mathbf{B}\cdot\mathbf{u}\right) }{6\pi^2v_F^2c\hbar^2} + \frac{e^2 \left(\mathbf{B}\cdot\bm{\omega}\right)}{24\pi^2 c v_F \hbar}
-\frac{e\mu \bm{\omega}^2}{6\pi^2v_F^3\hbar},\\
\label{app-current-J-matter}
\mathbf{J}_{\rm mat}+\mathbf{J}_{\rm mag} &=& -en\mathbf{u} -\frac{e\mu\mu_5 \bm{\omega}}{\pi^2v_F^2\hbar^2} +\frac{e^2\mu_5 \mathbf{B}}{2\pi^2\hbar^2c}
+\frac{e^2\mu_5 \left[\mathbf{E}\times\mathbf{u}\right]}{6\pi^2\hbar^2v_F^2}
+ \frac{e^2\left[\mathbf{E}\times\bm{\omega}\right]}{24\pi^2v_F\hbar}
+\frac{e\left[\mathbf{u}\times\bm{\nabla}\right]\mu\mu_5}{3\pi^2 v_F^2\hbar^2} -\frac{e \left[\bm{\nabla}\times\bm{\omega}\right] \mu}{12\pi^2 \hbar v_F}.
\end{eqnarray}
Note that the magneto-vortical correction $\propto \mathbf{B}\cdot\bm{\omega}$ to the charge density in the last expression is $6$ times smaller than in Ref.~[S14]. The origin of the discrepancy is not completely clear, but may indicate that the derivation of such terms from the CKT formalism is not completely reliable.

The chiral charge and current densities could be readily calculated by inserting the chirality operator $\lambda$ in Eqs.~(\ref{app-current-J0-def}), (\ref{app-current-J-def}), and (\ref{CKT-electric-current-magnetization-def}). The final
expressions for the nontopological parts of the chiral electric charge and current densities are
\begin{eqnarray}
\label{app-current-J50-matter}
\rho_{\rm mat,5} &=& -en_{5} -\frac{e^2\mu\left(\mathbf{B}\cdot\mathbf{u}\right) }{6\pi^2v_F^2c\hbar^2}  +\frac{e\mu_5 \bm{\omega}^2}{6\pi^2v_F^3\hbar},\\
\label{app-current-J5-matter}
\mathbf{J}_{\rm mat,5} + \mathbf{J}_{\rm mag,5}&=& -en_{5}\mathbf{u} -\frac{e\bm{\omega}}{2\pi^2v_F^2\hbar^2}\left(\mu^2+\mu_5^2+\frac{\pi^2T^2}{3}\right)
+\frac{e^2\mu \mathbf{B}}{2\pi^2\hbar^2c}
+\frac{e^2\mu \left[\mathbf{E}\times\mathbf{u}\right]}{6\pi^2\hbar^2v_F^2}  \nonumber\\
&+& \frac{e\left[\mathbf{u}\times\bm{\nabla}\right]}{6\pi^2 v_F^2\hbar^2} \left(\mu^2+\mu_5^2+\frac{\pi^2T^2}{3}\right) -\frac{e \left[\bm{\nabla}\times\bm{\omega}\right] \mu_5}{12\pi^2 \hbar v_F}.
\end{eqnarray}

\section{II. Linearized hydrodynamics equations for collective modes}
\label{sec:linearized}

In the study of collective modes, the deviations of the local thermodynamic parameters from
their global equilibrium values are small. Then, the use of linearized hydrodynamic equations
is sufficient. By looking for a solution in the
form of plane waves, i.e., $\delta\mu(\mathbf{r})=\delta\mu\, e^{-i\omega t+i\mathbf{k}\mathbf{r}}$
together with similar expressions for $\delta\mu_5(\mathbf{r})$, $\delta T(\mathbf{r})$, $\delta\mathbf{u}(\mathbf{r})$,
$\delta\mathbf{E}(\mathbf{r})$, and $\delta\mathbf{B}(\mathbf{r})$, we find the following
linearized form of the energy conservation relation (\ref{app-energy-der-result}):
\begin{eqnarray}
\label{ColEx-Energy-Fourier}
&&\omega \delta\epsilon
-(\epsilon+P)(\mathbf{k}\cdot\delta\mathbf{u}) -v_F\left(\mathbf{B}_0\cdot\mathbf{k}\right)\delta\sigma^{(\epsilon,B)}
-\sigma^{(\epsilon, u)} \left[
\left(\mathbf{B}_0\cdot \mathbf{k}\right)\left(\mathbf{B}_0\cdot\delta \mathbf{u}\right) - 2B_0^2\left(\mathbf{k}\cdot\delta \mathbf{u}\right)
\right]
=i\sigma^{(B)}\left(\mathbf{B}_0\cdot\delta\mathbf{E}\right),
\end{eqnarray}
where
\begin{equation}
\label{app-MHD-sigma-CKT-be}
\sigma^{(B)} = \frac{e^2\mu_5}{2\pi^2\hbar^2c}, \quad \sigma^{(\epsilon, B)}
= -\frac{e\mu\mu_5}{2\pi^2\hbar^2v_F c}, \quad \sigma^{(\epsilon, u)} = \frac{e^2v_F}{120\pi^2c^2\hbar}.
\end{equation}

The linearized Euler equation (\ref{app-Euler-der-result}) is
\begin{eqnarray}
\label{ColEx-Euler-Fourier}
&&\frac{\omega}{v_F}\left\{\frac{(\epsilon+P)}{v_F} \delta\mathbf{u}
+\mathbf{B}_0\delta\sigma^{(\epsilon,B)} + \sigma^{(\epsilon,B)}
\delta\mathbf{B} +i\frac{\hbar n_{5}\left[\mathbf{k}\times\delta\mathbf{u}\right]}{4} \right\} 
-\mathbf{k}\delta P+\frac{4 \sigma^{(\epsilon,B)}}{15v_F}
\left[ \mathbf{k}\left(\mathbf{B}_0\cdot\delta\mathbf{u}\right) +\left(\mathbf{B}_0\cdot\mathbf{k}\right)\delta\mathbf{u}
+\mathbf{B}_0\left(\mathbf{k}\cdot\delta\mathbf{u}\right)\right] \nonumber\\
&&+\frac{c}{3v_F}\sigma^{(\epsilon, B)}\left[\mathbf{k}\times\delta\mathbf{E}\right]
-5\sigma^{(\epsilon, u)}\mathbf{k}\left(\mathbf{B}_{0}\cdot\delta \mathbf{B}\right) 
-i\frac{\sigma^{(\epsilon, V)}}{5c}
\left\{(\mathbf{B}_0\cdot\mathbf{k})[\mathbf{k}\times\delta\mathbf{u}] +\mathbf{k}\left(\mathbf{B}_0\cdot[\mathbf{k}\times\delta\mathbf{u}]\right)\right\} \nonumber\\
&&=-ien\delta\mathbf{E} +\frac{i}{c}\left[\mathbf{B}_0\times \left(en\delta\mathbf{u}
- i\frac{\sigma^{(V)}}{3}[\mathbf{k}\times\delta\mathbf{u}] \right)\right] -i\frac{(\epsilon+P) \delta\mathbf{u}}{v_F^2\tau} +\frac{\hbar n_{5} [\mathbf{k}\times\delta\mathbf{u}]}{4v_F\tau}.
\end{eqnarray}
Here, $\sigma^{(V)}=-e\mu\mu_{5}/(\pi^2v_F^2\hbar^2)$,  $\sigma^{(\epsilon, V)} = -e\mu/(6\pi^2\hbar v_F)$, and the terms $\propto\left[\mathbf{k}\times\delta\mathbf{u}\right]$ stem from the vorticity effects. These
hydrodynamic equations should be also supplemented by
the Maxwell's equations as well as the continuity relations for the electric and chiral
charge densities.

\section{III. Useful formulas and relations}
\label{sec:App-ref}

In this section, we present useful formulas and relations used in the derivation of the Euler equation, the energy conservation relation, as well as the charge and current densities.
By making use of Eq.~(\ref{app-hydro-der-f0})
with $\epsilon_{p}=v_Fp$, it is straightforward to derive the following formulas:
\begin{eqnarray}
\int\frac{d^3p}{(2\pi)^3} p^{n-2}  f^{(0)}_{\lambda}
&=& -\frac{T^{n+1} \Gamma(n+1) }{2\pi^2 v_F^{n+1}}  \mbox{Li}_{n+1}\left(-e^{\mu_{\lambda}/T}\right),
\qquad n\geq 0,
\label{integral-3a} \\
\int\frac{d^3p}{(2\pi)^3} p^{n-2} \frac{\partial f^{(0)}_{\lambda}}{\partial \epsilon_{p}}
&=& \frac{T^{n} \Gamma(n+1) }{2\pi^2 v_F^{n+1}}  \mbox{Li}_{n}\left(-e^{\mu_{\lambda}/T}\right),
\qquad n\geq 0,
\label{integral-3b}
\end{eqnarray}
where $T\partial f^{(0)}_{\lambda}/\partial \epsilon_{p}
=-T\partial f^{(0)}_{\lambda}/\partial \mu_{\lambda}
=-e^{(\epsilon_{p}-\mu_{\lambda})/T}/[e^{(\epsilon_{p}-\mu_{\lambda})/T}+1]^2$ and $\mbox{Li}_{n}(x)$ is the polylogarithm function.
The polylogarithm functions for
$n=0,1$ are expressed through the elementary functions
\begin{eqnarray}
\mbox{Li}_{0}\left(-e^{x}\right) &=& -\frac{1}{1+e^{-x}}, \\
\mbox{Li}_{1}\left(-e^{x}\right) &=& -\ln{\left(1+e^{x}\right)}.
\label{App-polylog}
\end{eqnarray}

The following identities for the polylogarithm functions are useful when taking into account the holes (antiparticles) contributions:
\begin{eqnarray}
\mbox{Li}_{0} (-e^{x}) +\mbox{Li}_{0} (-e^{-x})   &=& -1,\\
\mbox{Li}_{1} (-e^{x}) -\mbox{Li}_{1} (-e^{-x})   &=& -x,\\
\mbox{Li}_{2} (-e^{x}) +\mbox{Li}_{2} (-e^{-x})   &=& -\frac{1}{2}\left(x^2+\frac{\pi^2}{3}\right),\\
\mbox{Li}_{3} (-e^{x}) - \mbox{Li}_{3} (-e^{-x})   &=&  -\frac{x}{6}\left(x^2+\pi^2\right),\\
\mbox{Li}_{4} (-e^{x}) + \mbox{Li}_{4} (-e^{-x})   &=&  -\frac{1}{4!}\left(x^4+2\pi^2x^2+\frac{7\pi^4}{15}\right).
\end{eqnarray}

By integrating over the angular coordinates, one can derive the following general relations:
\begin{eqnarray}
\int \frac{d^3 p}{(2\pi)^3}  \mathbf{p} \, f(p^2)&=& 0,\\
\int \frac{d^3 p}{(2\pi)^3}  \mathbf{p} (\mathbf{p}\cdot \mathbf{a}) f(p^2) &=& \frac{\mathbf{a}}{3}
\int \frac{d^3 p}{(2\pi)^3}  p^2 f(p^2),\\
\int \frac{d^3 p}{(2\pi)^3}  \mathbf{p} (\mathbf{p}\cdot \mathbf{a}) (\mathbf{p}\cdot \mathbf{b})  f(p^2) &=& 0,\\
\int \frac{d^3 p}{(2\pi)^3}  \mathbf{p} (\mathbf{p}\cdot \mathbf{a})(\mathbf{p}\cdot \mathbf{b})(\mathbf{p}\cdot \mathbf{c}) f(p^2) &=& \frac{1}{15}\left[\mathbf{a}(\mathbf{b}\cdot\mathbf{c})+\mathbf{b}(\mathbf{a}\cdot\mathbf{c})+\mathbf{c}(\mathbf{a}\cdot\mathbf{b})\right] \int \frac{d^3 p}{(2\pi)^3}  p^4 f(p^2).
\end{eqnarray}

\vspace{0.5cm}
\begin{center}
\noindent\rule{6cm}{1pt}
\end{center}
\vspace{0.5cm}

\begin{itemize}

\item[[S1\!\!]] E.~M.~Lifshitz and L.~P.~Pitaevskii,  {\sl Physical Kinetics} (Pergamon Press, New York, 1981).

\item[[S2\!\!]] K.~Huang,  {\sl Statistical Mechanics} (John Wiley and Sons, New York, 1987).

\item[[S3\!\!]] M.~A.~Stephanov and Y.~Yin,
  %``Chiral Kinetic Theory,''
  Phys.\ Rev.\ Lett.\  {\bf 109}, 162001 (2012).
  %doi:10.1103/PhysRevLett.109.162001
  %[arXiv:1207.0747 [hep-th]].

\item[[S4\!\!]] D.~T.~Son and N.~Yamamoto,
  %``Kinetic theory with Berry curvature from quantum field theories,''
  Phys.\ Rev.\ D {\bf 87}, 085016 (2013).
  %doi:10.1103/PhysRevD.87.085016
  %[arXiv:1210.8158 [hep-th]].
  %%CITATION = doi:10.1103/PhysRevD.87.085016;%%

\item[[S5\!\!]] M.~V.~Berry, Proc. R. Soc. A {\bf 392}, 45 (1984).
    %Quantal phase factors accompanying adiabatic changes

    %\bibitem{Chen:2014cla}
\item[[S6\!\!]]
  J.~Y.~Chen, D.~T.~Son, M.~A.~Stephanov, H.~U.~Yee, and Y.~Yin,
  %``Lorentz Invariance in Chiral Kinetic Theory,''
  Phys.\ Rev.\ Lett.\  {\bf 113}, 182302 (2014).
  %doi:10.1103/PhysRevLett.113.182302
  %[arXiv:1404.5963 [hep-th]].
  %%CITATION = doi:10.1103/PhysRevLett.113.182302;%%

%      \bibitem{Hidaka:2016yjf}
%\item[[S7\!\!]]
%  Y.~Hidaka, S.~Pu, and D.~L.~Yang,
  %``Relativistic Chiral Kinetic Theory from Quantum Field Theories,''
%  Phys.\ Rev.\ D {\bf 95}, 091901 (2017).
  %doi:10.1103/PhysRevD.95.091901
  %[arXiv:1612.04630 [hep-th]].

\item[[S7\!\!]]
    %\bibitem{Kharzeev:2007tn}
    D.~Kharzeev and A.~Zhitnitsky,
  %``Charge separation induced by P-odd bubbles in QCD matter,''
  Nucl.\ Phys.\ A {\bf 797}, 67 (2007).

\item[[S8\!\!]]
	%\bibitem{Kharzeev:2007jp}
D.~E.~Kharzeev, L.~D.~McLerran, and H.~J.~Warringa,
  %``The Effects of topological charge change in heavy ion collisions: 'Event by event P and CP violation',''
  Nucl.\ Phys.\ A {\bf 803}, 227 (2008).
  %[arXiv:0711.0950 [hep-ph]].
  %%CITATION = ARXIV:0711.0950;%%

\item[[S9\!\!]]
	%\bibitem{Fukushima:2008xe}
K.~Fukushima, D.~E.~Kharzeev, and H.~J.~Warringa,
  %``The Chiral Magnetic Effect,''
  Phys.\ Rev.\ D {\bf 78}, 074033 (2008).
  %%CITATION = ARXIV:0808.3382;%%

  %\bibitem{Bardeen}
   \item[[S10\!\!]]
  W.~A.~Bardeen, Phys. Rev. {\bf 184}, 1848 (1969);
 W.~A.~Bardeen and B.~Zumino,
  %``Consistent and Covariant Anomalies in Gauge and Gravitational Theories,''
  Nucl. Phys. B {\bf 244}, 421 (1984).
  %doi:10.1016/0550-3213(84)90322-5
  %%CITATION = doi:10.1016/0550-3213(84)90322-5;%%

	%\bibitem{Landsteiner:2013sja}
  \item[[S11\!\!]]
    K.~Landsteiner,
  %``Anomalous transport of Weyl fermions in Weyl semimetals,''
  Phys. Rev. B {\bf 89}, 075124 (2014).
  %doi:10.1103/PhysRevB.89.075124
  %[arXiv:1306.4932 [hep-th]].
  %%CITATION = doi:10.1103/PhysRevB.89.075124;%%

	%\bibitem{Landsteiner:2016}
  \item[[S12\!\!]]
    K.~Landsteiner,
 Acta Phys. Polonica B {\bf 47}, 2617 (2016).
  %``Notes on Anomaly Induced Transport,''
 % arXiv:1610.04413. % [hep-th].
  %%CITATION = ARXIV:1610.04413;%%

  %\bibitem{Gorbar:2017-Bardeen}
    \item[[S13\!\!]]
     E.~V.~Gorbar, V.~A.~Miransky, I.~A.~Shovkovy, and P.~O.~Sukhachov,
  %``Origin of the Bardeen-Zumino current in lattice models of Weyl semimetals,''
  Phys. Rev. B {\bf 96}, 085130 (2017).
  %arXiv:1706.02705.% [cond-mat.mes-hall].
  %%CITATION = ARXIV:1706.02705;%%

  \item[[S14\!\!]]
  K.~Hattori and Y.~Yin,
  %``Charge redistribution from anomalous magnetovorticity coupling,''
  Phys.\ Rev.\ Lett.\  {\bf 117}, 152002 (2016).
  %doi:10.1103/PhysRevLett.117.152002
  %[arXiv:1607.01513 [hep-th]].
  %%CITATION = doi:10.1103/PhysRevLett.117.152002;%%

\end{itemize}

%\end{widetext}